\documentclass[aip,graphicx]{revtex4-1}

\pdfoutput=1
\draft 

\usepackage{amsmath}
\usepackage{graphicx}
\usepackage{color}
\usepackage{bm}

\usepackage{soul}
\usepackage[normalem]{ulem}
\usepackage{multirow}
\usepackage{tabularx}
\usepackage[separate-uncertainty]{siunitx}

\begin{document}


\title{Characterization of a half-wave plate for cosmic microwave background circular polarization measurement with POLARBEAR} 



\author{T. Fujino\footnote{The author to whom correspondence may be addressed: \href{mailto:fugino-takuro-yk@ynu.jp}{email:fugino-takuro-yk@ynu.jp}}} 
\affiliation{Graduate School of Engineering Science, Yokohama National University, Yokohama, 240-8501, Japan}
\author{S. Takakura}
\affiliation{Department of Astrophysical and Planetary Sciences, University of Colorado Boulder, Boulder, CO 80309, USA} 
\author{Y. Chinone}
\affiliation{International Center for Quantum-field Measurement Systems for Studies of the Universe and Particles (QUP), High Energy Accelerator Research Organization (KEK), Tsukuba, Ibaraki 305-0801, Japan} 
\affiliation{Kavli Institute for the Physics and Mathematics of the Universe (Kavli IPMU, WPI), UTIAS, The University of Tokyo, Kashiwa, Chiba 277-8583, Japan} 
\author{M. Hasegawa}
\affiliation{Institute of Particle and Nuclear Studies (IPNS), High Energy Accelerator Research Organization (KEK), Tsukuba, Ibaraki 305-0801, Japan} 
\affiliation{International Center for Quantum-field Measurement Systems for Studies of the Universe and Particles (QUP), High Energy Accelerator Research Organization (KEK), Tsukuba, Ibaraki 305-0801, Japan} 
\affiliation{The Graduate University for Advanced Studies (SOKENDAI), Miura District, Kanagawa 240-0115, Hayama, Japan} 
\author{M. Hazumi}
\affiliation{International Center for Quantum-field Measurement Systems for Studies of the Universe and Particles (QUP), High Energy Accelerator Research Organization (KEK), Tsukuba, Ibaraki 305-0801, Japan} 
\affiliation{Kavli Institute for the Physics and Mathematics of the Universe (Kavli IPMU, WPI), UTIAS, The University of Tokyo, Kashiwa, Chiba 277-8583, Japan} 
\affiliation{Institute of Particle and Nuclear Studies (IPNS), High Energy Accelerator Research Organization (KEK), Tsukuba, Ibaraki 305-0801, Japan} 
\affiliation{Japan Aerospace Exploration Agency (JAXA), Institute of Space and Astronautical Science (ISAS), Sagamihara, Kanagawa 252-5210, Japan} 
\affiliation{The Graduate University for Advanced Studies (SOKENDAI), Miura District, Kanagawa 240-0115, Hayama, Japan} 
\author{N. Katayama}
\affiliation{Kavli Institute for the Physics and Mathematics of the Universe (Kavli IPMU, WPI), UTIAS, The University of Tokyo, Kashiwa, Chiba 277-8583, Japan} 
\author{A. T. Lee}
\affiliation{Department of Physics, University of California, Berkeley, Berkeley, CA 94720, USA} 
\affiliation{Physics Division, Lawrence Berkeley National Laboratory, Berkeley, CA 94720, USA} 
\affiliation{International Center for Quantum-field Measurement Systems for Studies of the Universe and Particles (QUP), High Energy Accelerator Research Organization (KEK), Tsukuba, Ibaraki 305-0801, Japan} 
\author{T. Matsumura}
\affiliation{Kavli Institute for the Physics and Mathematics of the Universe (Kavli IPMU, WPI), UTIAS, The University of Tokyo, Kashiwa, Chiba 277-8583, Japan} 
\author{Y. Minami}
\affiliation{Research Center for Nuclear Physics, Osaka University, Ibaraki, Osaka, 567-0047, Japan}
\author{H. Nishino}
\affiliation{The University of Tokyo, Tokyo, Research Center for the Early Universe, School of Science, 113-0033, Japan} 


\date{\today}

\begin{abstract}
A half-wave plate (HWP) is often used as a modulator to suppress systematic error 
in the measurements of cosmic microwave background (CMB) polarization.
A HWP can also be used to measure circular polarization (CP) through its optical leakage from CP to linear polarization.
The CP of the CMB is predicted
from various sources, such as interactions in the Universe and extension of the standard model.
Interaction with supernova remnants of population III stars is one of the brightest CP sources.
Thus, the observation of the CP of CMB is a new tool for searching for population III stars.
In this paper, we demonstrate the improved measurement of the leakage coefficient using the transmission measurement of an actual HWP in the laboratory.
We measured the transmittance of linearly polarized light through the HWP used in \textsc{Polarbear} in the frequency range of \SIrange{120}{160}{GHz}.
We evaluate properties of the HWP by fitting the data with a physical model using the Markov Chain Monte Carlo method.
We then estimate the band-averaged CP leakage coefficient using the physical model.
We find that the leakage coefficient strongly depends on the spectra of CP sources.
We thus calculate the maximum fractional leakage coefficient from CP to linear polarization as $0.133 \pm 0.009$ in the Rayleigh--Jeans spectrum.
The nonzero value shows that \textsc{Polarbear} has sensitivity to CP.
Additionally, because we use the bandpass of detectors installed in the telescope to calculate the band-averaged values, we also consider systematic effects in the experiment.
\end{abstract}


\maketitle 



\newcolumntype{C}{>{\centering\arraybackslash}p{20mm}}
\section{\label{intro}Introduction}
Measuring the polarization of the cosmic microwave background (CMB) is a powerful method of probing the physics of the early universe.
The linear polarization of the CMB comes from the quadrupole anisotropy at the last scattering of CMB photons.
There are two sources of quadrupole anisotropy, namely the density perturbation and primordial gravitational wave.
The density perturbation makes an even-parity linear polarization pattern called the $E$-mode in the CMB polarization whereas the primordial gravitational wave generated by cosmic inflation in the early universe makes an odd-parity linear polarization pattern called the $B$-mode as well as the $E$-mode.
The primordial $B$-mode polarization 
can be considered
a smoking gun for the inflation.
The scientific goal of many ongoing and future planed CMB experiments is finding the signal in the linear polarization.

CMB photons can also have circular polarization (CP).
Although the CP of the CMB is not predicted by the standard models of cosmology, the so-called $\Lambda$ Cold Dark Matter ($\Lambda$CDM) model,
there are mechanisms that can generate the CP of the CMB photons during their propagation from the last scattering to an observer today.
Examples are Faraday conversion (FC) by the magnetic fields of galaxy clusters\cite{circ_galaxy},
FC by relativistic plasma remnants of Population III stars \cite{circ_popIII},
scattering from the cosmic neutrino background (C$\nu$B) \cite{circ_CnB}, and photon-photon-scattering \cite{circ_photon_photon}.
There also are predictions of the CP of the CMB due to the extension of the standard model of the cosmology and particle physics, such as cosmological pseudoscalar fields\cite{circ_pseudo_field}.

Some CMB polarization experiments have set an experimental constraint on the angular power spectrum of CP.
The optics of the Cosmology Large Angular Scale Surveyor (CLASS) have sensitivity to CP 
because of the use of a
variable-delay polarization modulator \cite{CLASS_VPM}.
They placed the best upper limit on the CP power spectrum at the degree scale\cite{circ_CLASS_result}.
SPIDER, a balloon-borne telescope designed to search for the $B$-mode linear polarization of the CMB, placed a limit on the CP power spectrum even though the telescope was not designed to be sensitive to the CP.
The SPIDER project utilized the imperfection of the half-wave plate (HWP) modulator~\cite{circ_SPIDER_result}.
An ideal HWP works as a retarder of phase $\pi$, inverting the electric field of one axis of the incident linearly polarized light.
The outgoing light is thus linearly polarized at a different angle from the incoming light.
In practice, however, a HWP comprising single-layer birefringent material only satisfies this condition at a target frequency.
Outside the target frequency range within the observational frequency band, the retardance is no longer $\pi$,
and a coupling between linear polarization and CP is thus created.
In this case,
a part of the incident CP leaks to linear polarization.
Thus, a CMB telescope, which is designed to detect a linearly polarized signal, gains sensitivity to CP
as long as the properties of its HWP are well characterized.

To utilize this small leakage,
we need to know the amount of leakage precisely by calibration with known CP light or characterization from measured optical parameters.
Because we do not have a calibrator for CP in \textsc{Polarbear}, 
in this paper,
we choose to characterize the leakage of the HWP.
Although there have been papers on CMB experiments using a HWP as a polarization modulator\cite{hwp_chara_MAXIPOL, hwp_chara_PB2, ABS_results}, they mainly reported only the efficiency of the linear polarization and did not include the leakage between CP and linear polarization because they aimed to measure the linear polarization of the CMB.
Some papers\cite{hwp_chara_spider, hwp_chara_LB} reported the leakage between the CP and linear polarization using Mueller matrix formalism; however, these were simulation-only results or proto-type results, not evaluation results of a HWP for actual CMB observation.
A study on CP by SPIDER\cite{circ_SPIDER_result} used
the leakage estimated from the design values of the HWP and calculated an upper limit on the angular power spectrum of the CMB CP.

In the present paper, we report the characterization of a HWP installed in the \textsc{Polarbear} telescope.
\textsc{Polarbear} is a CMB experiment that began in January 2012.
From May 2014 to December 2016, \textsc{Polarbear} performed large-angular-scale observations using the HWP, which was continuously rotated at \SI{2}{Hz}.
With these data, we measured the CMB $B$-mode power spectrum over a multipole range of $50 < \ell < 600$ and put a limit on the tensor-to-scalar ratio, $r < 0.33$\cite{reanalysis}.
We evaluate the HWP using transmission spectrum taken in the laboratory in 2014 before its installation in the \textsc{Polarbear} telescope.
We determine physical parameters of the HWP by fitting the data with a theoretical transmission model and then estimate the leakage between the CP and linear polarization using this model.
Through measurement of the fringe pattern of the HWP, we determine the thickness and the difference in the refractive index between the ordinary and extra-ordinary axes of sapphire precisely.

Section~\ref{sec2} presents an overview of the CMB polarization observation with the continuously rotating HWP.
Section~\ref{sec3} explains the \textsc{Polarbear} experiment.
Section~\ref{sec4} details the characterization of the \textsc{Polarbear} HWP in the laboratory.
We also present the results of the transmittance of the HWP for linearly polarized light.
In section~\ref{sec5}, we explain a method of estimating the HWP leakage between CP and linear polarization and present the results.
We discuss systematic uncertainty of the leakage in section~\ref{sec6}.
We also discuss the prospects of CP measurements using \textsc{Polarbear}.

\section{\label{sec2}Polarization Observation with a HWP}
A HWP is an optical device that creates an optical path difference of half of the wavelength.
CMB experiments widely use HWPs made from birefringent materials.
A HWP converts linearly polarized incident light with angle $\alpha_{in}$ to linearly polarized light with angle $2 \theta_h-\alpha_{in}$, where $\theta_{h}$ is the angle of the fast axis of the HWP.
In \textsc{Polarbear}, we rotate the HWP continuously 
to
separate the linear polarization signal from the unpolarized signal in the frequency domain 
to reduce
systematic uncertainties, which are generated in the instruments after the HWP and by the detector pair difference, and low-frequency noise.
In this section, we explain the optical model of the HWP and how the HWP modulates polarization signals.

\subsection{HWP modeling}
\label{sec2-1}
\begin{figure}[tb]
    \centering
    \includegraphics[width=10cm]{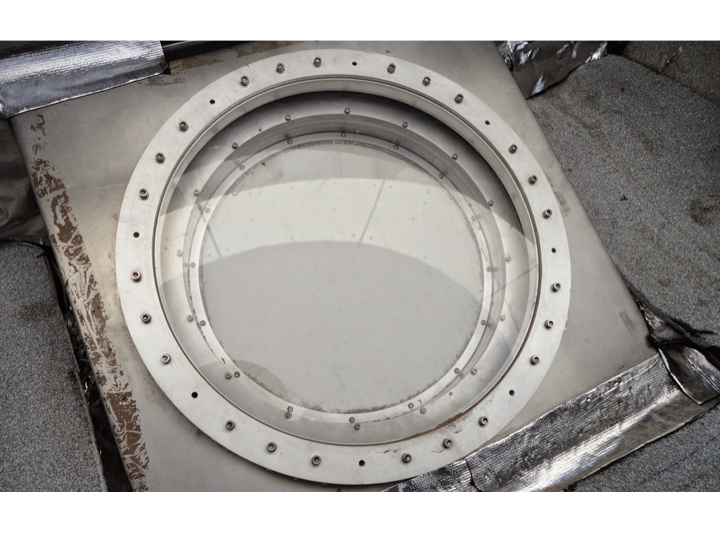}
    \caption{Photograph of the HWP in the \textsc{Polarbear} experiment.}
    \label{fig:hwp_photo}
\end{figure}

The HWP used in \textsc{Polarbear} comprises
a \SI{28} {cm} diameter \SI{3.1}{mm} thick single sapphire as birefringent material,
which is sandwiched between two anti-reflection coating layers of \SI{0.254} {mm} thick Duroid 6002.
They are attached with a glue layer comprising \SI{0.038} {mm}-thick polyethylene (LDPE).
The thickness of the birefringent material is determined so that the optical path difference between the slow axis and fast axis is half of the wavelength.
The HWP is shown in Figure~\ref{fig:hwp_photo} and the design values of the HWP are given in Table~\ref{tab:hwp_parameter}. 

\begin{table}[tb]
    \centering
    \caption{Design values of the HWP used in the \textsc{Polarbear} experiment. The first and second rows give values for the ordinary and extra-ordinary axes of the sapphire, respectively. The refractive indices and loss tangents of the components come from V. Parshin (1994)\cite{sapphire_para}, RT/duroid (2020)\cite{DUROID6002}, and J. W. Lamb (1996)\cite{LDPE_para}.}
    \begin{tabular}{cccc}
        \hline \hline
        & thickness & refractive index & loss tangent \\
        \hline
        Sapphire (o-axis) & \SI{3.1 +- 0.1}{mm} & \SI{3.068 +- 0.003}{} & \SI{2.30e-4}{} \\
        Sapphire (e-axis) & \SI{3.1 +- 0.1}{mm} & \SI{3.402 +- 0.003}{} & \SI{1.25e-4}{} \\
        Duroid & \SI{0.254}{mm} & \SI{1.715 +- 0.012}{} & \SI{12.0e-4}{} \\
        LDPE & \SI{0.038}{mm} & \SI{1.514 +- 0.010}{} & \SI{{\sim} 5.0e-4}{} \\
        \hline \hline
    \end{tabular}
    \label{tab:hwp_parameter}
\end{table}

To express the polarization state of light, we introduce the Stokes vector as
\begin{equation}
    \label{eq:stokes}
    P = \langle \bm{E E^{\dagger}} \rangle = I \sigma_0 + Q \sigma_3 + U \sigma_1 + V \sigma_2,
\end{equation}
where $\bm{E}$, $\bm{E}^\dagger$, $\sigma_{x}$, and the angled brackets denote the vector of complex electric fields, its complex conjugate transposition, Pauli matrices, and time averaging, respectively.
The elements of the Stokes vector, $I$, $Q$, $U$, and $V$, are respectively the intensity, linear polarization amplitude on the axes, amplitude of 45-degree-tilted linear polarization, and amplitude of CP.

We also introduce the Mueller matrix, which represents the conversion of the Stokes parameter by each optical element.
A Mueller matrix $M$ is expressed using the Jones matrix $J$ as
\begin{equation}
    \label{eq:jones_to_mueller}
    M_{ij} = \frac{1}{2} \mathrm{Tr}(\sigma_i J \sigma_j J^{\dagger}),
\end{equation}
where $i, j$ are indices of the matrix.
The Jones matrix is a $2 \times 2$ matrix expressing the complex transmission of the electric field of each axis,
\begin{equation}
    \label{eq:jones_def}
    \begin{pmatrix}
        E_{x, \mathrm{out}} \\
        E_{y, \mathrm{out}}
    \end{pmatrix}
    = J
    \begin{pmatrix}
        E_{x, \mathrm{in}} \\
        E_{y, \mathrm{in}}
    \end{pmatrix},
\end{equation}
where $E_{x, \mathrm{in}}, E_{y, \mathrm{in}}, E_{x, \mathrm{out}}, E_{y, \mathrm{out}}$ are components of incoming and outgoing complex electric fields.
In the case of HWP, since it works as a retarder, we can express $J$ by retardance $\delta$, transmittance of x-axis and y-axis $a(\nu), b(\nu)$, and coupling between axes $\epsilon_1(\nu), \epsilon_2(\nu)$ as
\begin{equation}
    \label{eq:jones_matrix}
    J =
    \begin{pmatrix}
        a(\nu) &\epsilon_1(\nu) \\
        \epsilon_2(\nu) &b(\nu) e^{i \delta(\nu)}
    \end{pmatrix}.
\end{equation}
In this expression, we assume that the $x$-axis is aligned with the ordinary axis of the HWP.

Because the HWP of \textsc{Polarbear} is made from a single-layer birefringent material in which the coupling of
$\epsilon_1(\nu)$ and $\epsilon_2(\nu)$ vanishes,
we neglect $\epsilon_1(\nu)$ and $\epsilon_2(\nu)$ hereafter.
Therefore, the Mueller matrix of the \textsc{Polarbear} HWP is expressed as
\begin{equation}
    \label{eq:mueller_matrix}
    M_{\mathrm{HWP}} = 
    \begin{pmatrix}
        T &\rho &0 &0 \\
        \rho &T &0 &0 \\
        0 &0 &c &-s \\
        0 &0 &s &c
    \end{pmatrix} = 
    \begin{pmatrix}
        a^2+b^2 &a^2-b^2 &0 &0 \\
        a^2-b^2 &a^2+b^2 &0 &0 \\
        0 &0 &ab \cos\delta &-ab \sin\delta \\
        0 &0 &ab \sin\delta &ab \cos\delta \\
    \end{pmatrix}.
\end{equation}
In this matrix, $T$, $\rho$, $c$, and $s$, denote the transmittance, differential transmittance between the two HWP axes, polarization efficiency, and coupling between linear polarization and CP states, respectively.
The retardance $\delta$ is the phase difference caused by this HWP as $\delta = 2 \pi (n_s-n_f) d \nu /v_c$, where $n_f$ and $n_s$ are the indices of the fast and slow axes, $d$ is the thickness of the birefringent material, $\nu$ is the electromagnetic frequency of the incoming radiation, and $v_c$ is the speed of light.
In the case of an ideal HWP,
these values are
$T=1, c=-1,$ and $\rho=s=0$.

We follow T. Essinger-Hileman (2013) \cite{HWP_modeling} in calculating the Mueller matrix of the HWP.
In this method, the Mueller matrix of a stack of isotropic and birefringent material layers is calculated using a generalized transfer matrix, which solves the boundary conditions of the electric and magnetic fields of the transmitted, reflected, and absorbed waves.
In this paper, we consider that the HWP comprises a sapphire and an anti-reflective (AR) coating (Duroid) with glue (LDPE) on both sides of the sapphire.
We use the thickness, refractive index, and loss tangent of each layer of the HWP to calculate the Mueller matrix of the HWP.
In this calculation, we assume that the thickness of the AR coating is the same on the two sides of the sapphire.
We also assume that the HWP is in air with a refractive index of 1 and that light enters the HWP vertically.
We discuss the effect of the case that light enters the HWP non-vertically, like the \textsc{Polarbear} telescope, in section \ref{sec6-1}.

\subsection{Polarization measurement with the HWP}
\label{sec2-2}
When a detector sensitive to a single linear polarization observes the sky through a HWP continuously rotating at an angular velocity of $\omega_h$, the observed quantity is the integral of the signal spectrum over the observational bandwidth: 
\begin{equation}
    \bar{d} = \int_0^\infty W(\nu) d(\nu) d\nu,
\end{equation}
where $d(\nu)$ is the detector signal at each frequency and $W(\nu)$ is the window function for the spectral band-shape.
For the incident signal with a Stokes vector of $(I(\nu), Q(\nu), U(\nu), V(\nu))$, the signal is derived as
\begin{equation}
    \label{eq:mod_signal}
    \begin{split}
        d(\nu) =\ &
        \frac{1}{2} V_{\mathrm{det}}
        M_{\mathrm{rot}}(-2\omega_h t) M_{\mathrm{HWP}}(\nu) M_{\mathrm{rot}}(2\omega_h t)
        \left( I(\nu),\ Q(\nu),\ U(\nu),\ V(\nu) \right)^{T} \\
        = \ & d_0(\nu) + d_2(\nu) + d_4(\nu).
    \end{split}
\end{equation}
Here,
$V_{\mathrm{det}} = ( 1,\ \cos(2\theta_{\mathrm{det}}),\ \sin(2\theta_{\mathrm{det}}),\ 0 )$ is the vector of the detector, $\theta_{\mathrm{det}}$ is the angle of the detector, and $M_{\mathrm{rot}}$ is the Mueller matrix of the coordinate rotation:
\begin{equation}
    \label{eq:rot_matrix}
    M_{\mathrm{rot}}(\theta) = 
    \begin{pmatrix}
        1 &0 &0 &0 \\
        0 & \cos(\theta) &\sin(\theta) &0 \\
        0 &-\sin(\theta) &\cos(\theta) &0 \\
        0 &0 &0 &1
    \end{pmatrix}.
\end{equation}
$d_0(\nu)$, $d_2(\nu)$, and $d_4(\nu)$ are respectively the zeroth, second, and fourth harmonics signals of the HWP rotation frequency,
\begin{align}
    \label{eq:mod_signal0}
    d_0(\nu) =& \frac{1}{2} \left( I(\nu)T(\nu) + Q(\nu) \frac{T(\nu)+c(\nu)}{2} \cos(2\theta_{\mathrm{det}}) - U(\nu) \frac{T(\nu)+c(\nu)}{2} \sin(2\theta_{\mathrm{det}}) \right), \\
    \label{eq:mod_signal2}
    \begin{split}
        d_2(\nu) =& \frac{1}{2} \left( I(\nu) \rho(\nu) \cos(2 \omega_h t - 2\theta_{\mathrm{det}}) +
        Q(\nu)  \rho(\nu) \cos(2 \omega_h t) - U(\nu) \rho(\nu) \sin(2 \omega_h t) \right. \\
        & \left. + V(\nu) s(\nu) \sin(2\omega_h t - 2\theta_{\mathrm{det}}) \right),
    \end{split}
    \\
    \label{eq:mod_signal4}
    d_4(\nu) =& \frac{1}{2} \left( Q(\nu) \frac{T(\nu)-c(\nu)}{2} \cos(4 \omega_h t - 2\theta_{\mathrm{det}})
    - U(\nu) \frac{T(\nu)-c(\nu)}{2} \sin(4 \omega_h t - 2\theta_{\mathrm{det}}) \right).
\end{align}
Equation~\eqref{eq:mod_signal4} shows that the $Q$ and $U$ signals are modulated into the fourth harmonic signal.
In a previous linear polarization observation\cite{large_patch}, we evaluated the polarization efficiency, which is shown as $(T-c)/2$ in Eq~\eqref{eq:mod_signal4}, by observation of the Crab Nebula and physical optics simulation.
Equation~\eqref{eq:mod_signal2} shows that the CP component $V$ is in the second harmonic signal.

Unlike $T(\nu)$ and $(T(\nu) - c(\nu))/2$, which are $\sim 1$ over the observational bandwidth, the other leakage coefficients, $(T(\nu) + c(\nu))/2$, $\rho(\nu)$, and $s(\nu)$, vary within the band.
We therefore use the band average of the leakage coefficient for the effective leakage coefficient:
\begin{equation}
    \label{eq:band_averaged}
    \bar{s} \equiv \frac{\int s(\nu) V(\nu) W(\nu) d\nu}{\int V(\nu) W(\nu) d\nu},
\end{equation}
where $V(\nu)$ is the source spectra.
If this $\bar{s}$ is nonzero, the detector has sensitivity to the incident CP signal.

From the modulated detector timestream, we extract each harmonic signal through demodulation using the recorded angle of the HWP.\cite{CRHWP2017}
However, unlike the fourth harmonic signal, the second harmonic signal contains not only the CP signal
but
also the intensity and linear polarization signals.
We thus need to eliminate these contaminations.
We can subtract the linear polarization component
considering the correlation with the second and fourth harmonic signals because the linear polarization signal is simultaneously present in the second and fourth harmonics signals.
The zeroth harmonic signal is available for the removal of intensity.
The coefficient of the linear polarization in the zeroth signal $T(\nu)+c(\nu) \ll 1$, and the intensity signal is thus dominant in the zeroth harmonic.
We thus subtract the intensity component from the correlation between the zeroth and second harmonic signals.

\section{POLARBEAR Experiment}
\label{sec3}
\begin{figure}
        \centering
        \includegraphics[width=7cm]{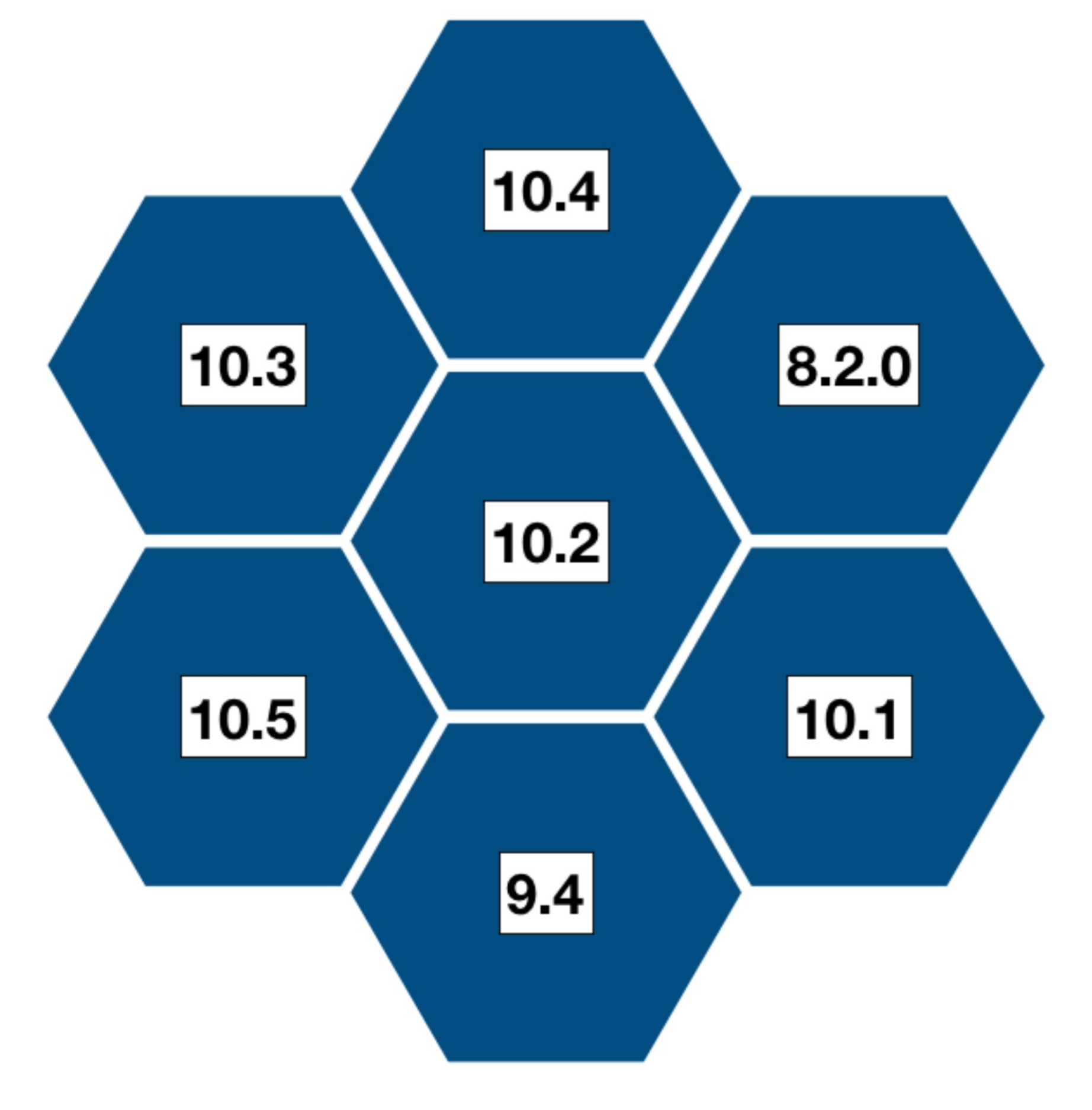}
    \caption{Layout of detector wafers on the focal plane of \textsc{Polarbear}\cite{FTScharacterization}. There are seven wafers on the \textsc{Polarbear} focal plane.
    Each wafer is labeled by the number of versions, and these numbers correspond to other \textsc{Polarbear} papers.
    Six wafers have silicon lenslets, and one wafer, which is labeled 8.2.0, has alumina lenslets.
    Reproduced from F. Matsuda et. al., "The POLARBEAR Fourier transform spectrometer calibrator and spectroscopic characterization of the POLARBEAR instrument”, Review of Scientific Instruments 90, 115115 (2019), with the permission of AIP Publishing.}
    \label{fig:pb_wafers}
\end{figure}

\textsc{Polarbear} is a CMB experiment conducted at the James Ax Observatory, which is at an altitude of \SI{5190}{m} in the Atacama Desert, Chile.
\textsc{Polarbear} searched for both degree-scale and sub-degree-scale $B$-mode polarization signals originating from inflationary gravitational waves and the weak gravitational lensing effect, respectively.
The Huan Tran Telescope, which is equipped with the \textsc{Polarbear} receiver, has an off-axis Gregorian optics configuration with a \SI{2.5}{m}-diameter primary mirror and secondary mirror.
The \textsc{Polarbear} receiver has seven wafers on the focal plane (see Figure~\ref{fig:pb_wafers}).
Each wafer is mounted with 182 detectors, and there is thus a total of 1274 detectors on the focal plane.
The detector is sensitive to the frequency band centered at \SI{150}{GHz} with a fractional band width of approximately \SI{30}{\%}.
The actual observation band is typically from \SIrange{128.0}{159.8}{GHz}.
See Table \ref{tab:wafer_bandpass} for details.
We measured the band-pass window function at the site with a Fourier transform spectrometer (FTS)
in April 2014. In this measurement, a Martin-Puplett interferometer is mounted on the telescope.
Thanks to high throughput, optimized optical coupling to the \textsc{Polarbear} optics using a custom designed output parabolic mirror, and a continuously modulated output polarizer, it measured the band-pass window function at a high signal-to-noise ratio.
In this paper, we use the wafer-averaged values of this measurement.
See F. Matsuda et al. (2019)\cite{FTScharacterization} for details.
\textsc{Polarbear} began observations in January 2012. We placed a continuously rotating HWP between the primary and secondary mirrors
to measure the degree-scale $B$-mode polarization from May 2014\cite{CRHWP2017}.
Note that the HWP was not installed during the FTS measurement.

\section{Laboratory Measurement}
\label{sec4}
We characterized the optical properties of the HWP
installed in the \textsc{Polarbear} telescope. 
The characterization was conducted in a laboratory environment prior to the deployment of the HWP to the \textsc{Polarbear} telescope. 

\subsection{Measurement System}
\label{sec4-1}
\begin{figure}[tbp]
    \centering
    \includegraphics[width=16cm]{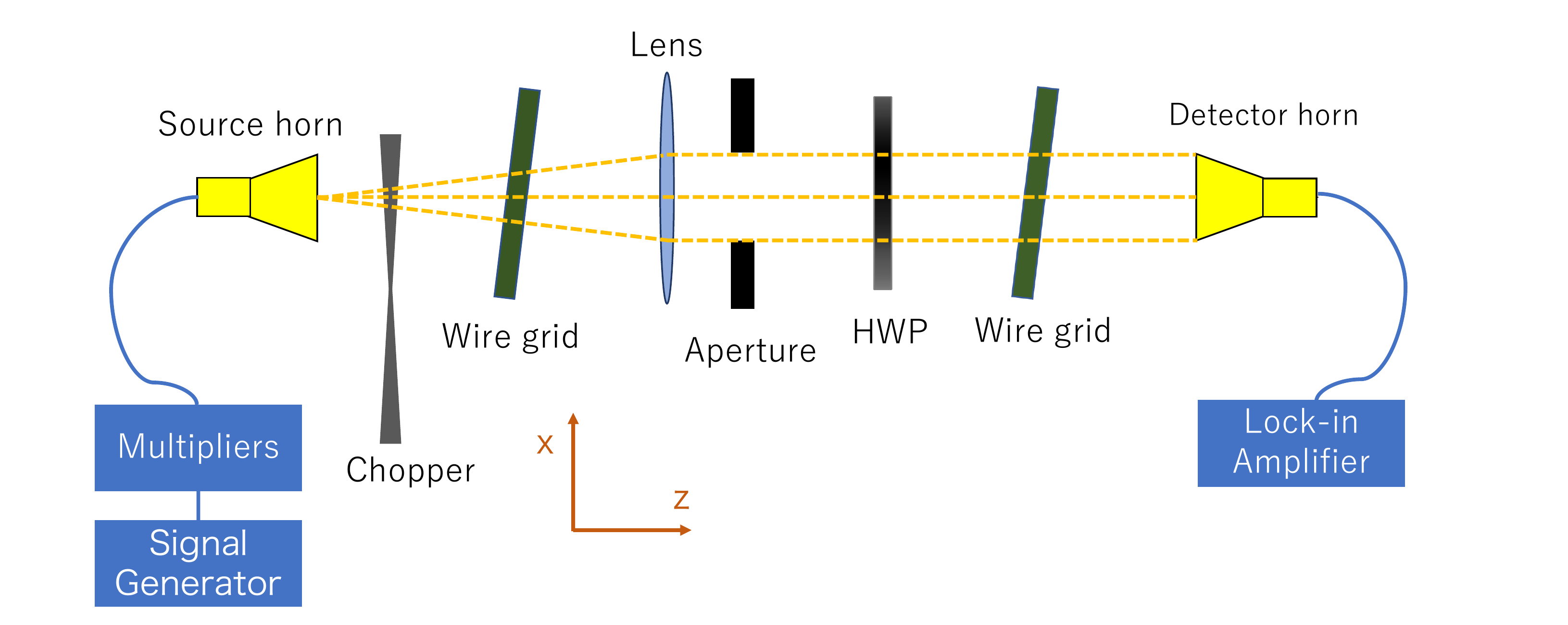}
    \caption{Setup of the transmission measurement of the HWP.
    The signal from 108 GHz to 162 GHz is emitted from the source horn at the left of the figure.
    The HWP is mounted on a holder that rotates about the z-axis.
    There are two wire grids before and after the HWP to define the incoming and outgoing polarized signal.
    The detector is set on a linear stage to calibrate the effect of standing waves.
    }
    \label{fig:lab_test}
\end{figure}

Figure~\ref{fig:lab_test} shows the measurement setup for characterizing the \textsc{Polarbear} HWP.
A millimeter-wave source signal is generated by a continuous wave generator (12 to 18~GHz) with a $\times~9$ multiplier (QUINSTAR QMM-E0FB00090) covering a frequency range from 108 to 162~GHz.
The source signal is emitted through the waveguide of WR6 with a pyramidal feed horn
(SGH-06-RP000).
After the feed horn, we place an optical chopper to chop the signal at \SI{11}{Hz}.
We insert a wire grid along the optical path to define the polarization angle. 
This wire grid is placed on a fixed folder and placed obliquely to the optical path to prevent standing waves.
The polarized signal is collimated by a lens before the aperture. 
The lens is composed of Rexolite with a sub-wave grading AR coating.
The HWP is mounted on a holder that rotates about the z-axis. 
The HWP is placed normal to the incident radiation. 
This rotational mechanism is controlled by a stepping motor.
In our measurement, the HWP is rotationally stepped in intervals of 6 degrees.
The angle of the fast axis of the HWP is originally unknown, and we thus calculate this angle in the data analysis.
Behind the HWP, another wire grid is inserted to determine the outgoing polarized signal.
This wire grid is also placed obliquely to the optical path, and placed on a fixed folder with the same angle of the other wire grid.
The detector is a diode detector that is sensitive to the input millimeter wave.
The detector is set on a linear stage to calibrate the effect of standing waves and moves 1.9~mm along the z-axis.
The detected signal is read by a lock-in amplifier with a reference signal from the chopper.

\subsection{Methods}
\label{sec4-2}
The following measurement method was adopted using the setup in Figure~\ref{fig:lab_test}. 
We start the measurement by setting the frequency of the signal generator to 12 GHz (the output frequency of the multiplier is 108 GHz) and the rotation angle of the HWP to 0 degrees.
In this measurement, we take data by changing the detector position along the z-axis from 0 to 1.9~mm.
After the measurement at 0 degrees and 12 GHz, we step the rotation angle of the stepping motor to 6 degrees and repeat the measurement.
After making measurements from 0 to 354 degrees, we step the frequency of the signal generator by 0.1 GHz (output 0.9 GHz).
We repeat the measurements until making measurements at 18 GHz.
That is, the maximum frequency of the measurement is 162 GHz.

We also measure the transmittance of the linear polarization by taking the ratio of the output signal when the HWP is inserted into the optical system and the output signal when the HWP is not inserted.
These measurements are performed after the angle of the fast axis of the HWP is determined by the previous measurement.

\subsection{Analysis}
\label{sec4-3}
In the analysis, we first remove the effect of the standing wave from the signal.
For each frequency and rotation angle, we plot the signal as a function of the linear stage position.
We then fit the signal with a sinusoidal wave, a drift, and a constant component and extract the constant component as the signal for this frequency and rotation angle.
See appendix \ref{app:standing_wave} for the detail.

\begin{figure}[tbp]
    \centering
    \includegraphics[width=13cm]{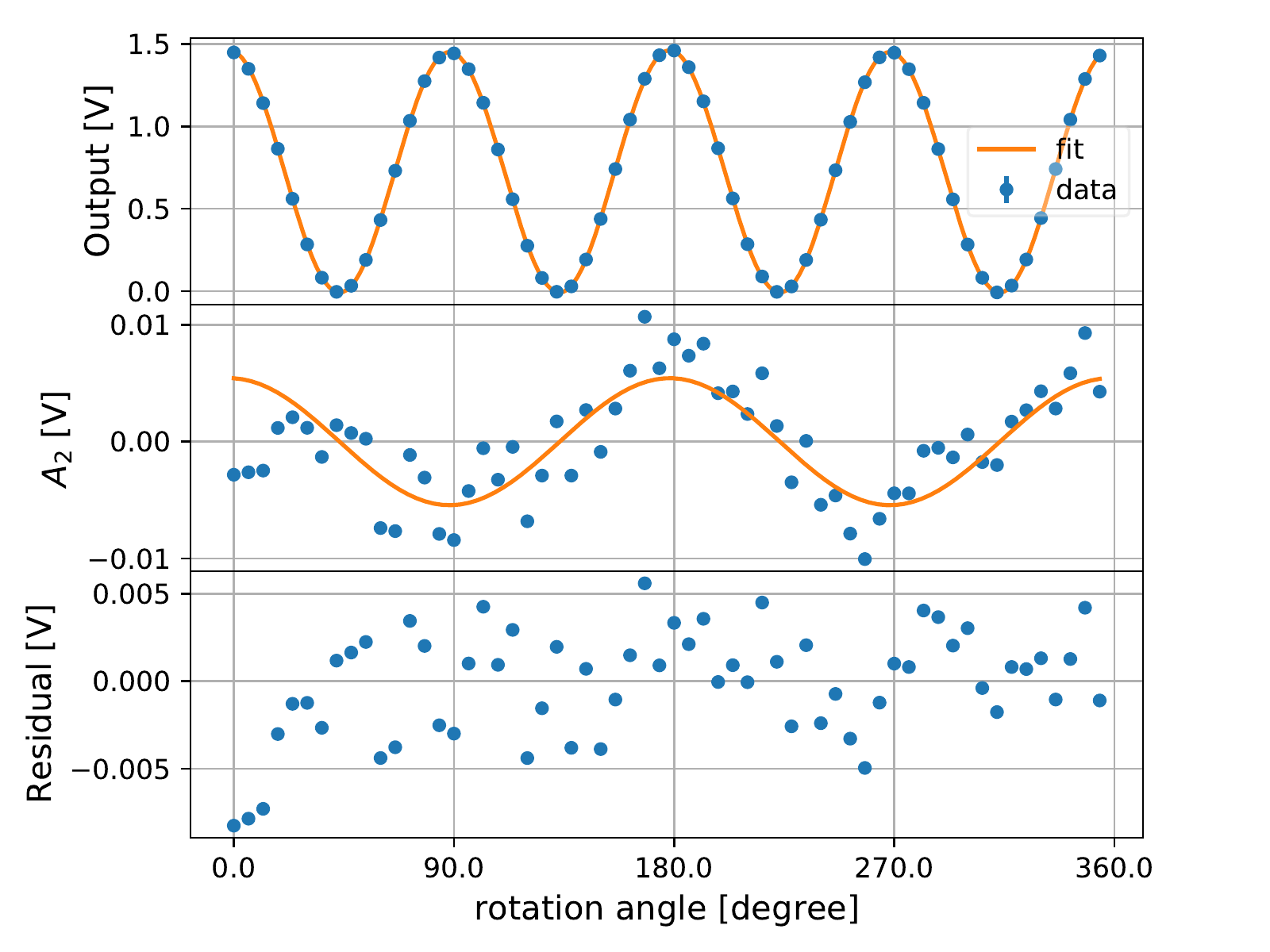}
    \caption{(Top) Example of the output signal for each rotation angle at an incident frequency of 143.1 GHz. The blue points are the data, and the orange line shows the fit with Eq.~\eqref{eq:data_fitmodel}. (Middle) The second harmonic component of the same data obtained by subtracting $A_0 + A_4 \cos(4\theta_h + 4\phi_4)$ of Eq.~\eqref{eq:data_fitmodel}. (Bottom) Residual of the fit.}
    \label{fig:raw_data}
\end{figure}

Then, given the above setup and the measurement procedure, the data are modeled as 
\begin{equation}
    \label{eq:sig_lab}
    \begin{split}
        d_m(\nu, \theta_h) = &G(\nu) \left( 1,\ 1,\ 0,\ 0 \right)
        M_{\mathrm{rot}}(-2\theta_h) M_{\mathrm{HWP}}(\nu) M_{\mathrm{rot}}(2\theta_h)
        \left( 1,\ 1,\ 0,\ 0 \right)^{T} + d_\mathrm{offset} \\
        = &G(\nu)\frac{3 T(\nu)+c(\nu)}{2} + 2G(\nu) \rho(\nu) \cos(2 \theta_h) \\
        &+ G(\nu)\frac{T(\nu)-c(\nu)}{2} \cos(4 \theta_h) + d_\mathrm{offset},
    \end{split}
\end{equation}
where $\theta_h$ is the rotation angle of the HWP, $G(\nu)$ is the gain of the system, and $d_\mathrm{offset}$ is the offset of the lock-in amplifier.
We find that some measurements are negative and assume that this is due to the offset of the lock-in amplifier.
We thus include $d_\mathrm{offset}$ as a parameter.
Given this model, we fit the data using the equation
\begin{equation}
    d = A_0 + A_2 \cos{(2\theta_h+2\phi_2)} + A_4 \cos{(4\theta_h+4\phi_4)}.
    \label{eq:data_fitmodel}
\end{equation}
Within the model in section \ref{sec2-2}, we can explain all data measured by this method by 0th, 2nd, and 4th harmonics.
Figure~\ref{fig:raw_data} shows a typical modulated response curve when the HWP rotates about the z-axis.
The top panel shows the modulated power as a function of the HWP angle at 143~GHz. 
The points are the measurement data, and the curve is the fit using Eq.~\eqref{eq:data_fitmodel}.
The middle panel shows the second harmonic component obtained by computing $d-(A_0+A_4 \cos{(4\theta_h+4\phi_4)})$ in Eq.~\eqref{eq:data_fitmodel}. 
The bottom panel shows the residual of the fit. 
Here, the peaks of the top panel correspond to the angles where the fast or slow axis of the HWP becomes parallel to the incident polarization.
There is a small difference in the transmission between the fast and slow axes owing to the difference in the refractive index, which is seen as the second harmonic component in the middle panel.
The signal becomes almost zero at the middle of the peaks.
This means that the input polarization is efficiently rotated to the orthogonal polarization, and the leakage to the CP is small.

We then relate the obtained amplitudes, $A_0$, $A_2$, and $A_4$, to the components of the Mueller matrix of the HWP.
By comparing Eqs~\eqref{eq:sig_lab} and \eqref{eq:data_fitmodel}, we obtain
\begin{equation}
    \label{eq:connections}
    \begin{split}
        \frac{\rho(\nu)}{T(\nu)} + \Delta_1(\nu) &= \frac{A_2(\nu)}{A_0(\nu)+A_4(\nu)}, \ \ 
        \frac{c(\nu)}{T(\nu)} + \Delta_2(\nu) = \frac{A_0(\nu)-3A_4(\nu)}{A_0(\nu)+A_4(\nu)}, \\
    \end{split}
\end{equation}
where $\Delta_1(\nu)$ and $\Delta_2(\nu)$ denote the effect of the offset of the lock-in amplifier.
These values are calculated using $d_\mathrm{offset}$:
\begin{align}
    \label{offset}
    \Delta_1(\nu) &= \frac{-A_2(\nu) d_\mathrm{offset}}{((A_0(\nu)+A_4(\nu))(A_0(\nu)-d_\mathrm{offset}+A_4(\nu))}, \\
    \Delta_2(\nu) &= \frac{4A_0(\nu) d_\mathrm{offset}}{((A_0(\nu)+A_4(\nu))(A_0(\nu)-d_\mathrm{offset}+A_4(\nu))}.
\end{align}
In this calculation, we assume the $d_\mathrm{offset}$ is constant for all frequencies.

We also calculate the transmittance of the linear polarization $T+\rho$ from the ratio of the observations made with a HWP rotation angle $\theta_h=0$ and observations made without the HWP as
\begin{equation}
    \label{eq:connections2}
    T(\nu)+\rho(\nu) = \frac{d_m (\theta_h=0)(\nu)}{d_\mathrm{no HWP}(\nu)}.
\end{equation}

Finally, we fit the spectra obtained from Eqs.~\eqref{eq:connections} and \eqref{eq:connections2} with the model of the Mueller matrix of the HWP described in section~\ref{sec2-1} using the Markov Chain Monte Carlo (MCMC) method with the thickness, refractive index, and loss tangent of each layer of the HWP as input parameters.
As the prior distribution of the HWP model, we use the design values given in Table~\ref{tab:hwp_parameter}.
To avoid negative values, we assume the exponential distribution as the prior distribution of loss tangents.
We assume Gaussian distributions for the prior distributions of other parameters.
Additionally, we include $d_\mathrm{offset}$ as the input parameter of the MCMC fitting.
The prior distribution of $d_\mathrm{offset}$ is a uniform distribution from -0.05 to 0.05 because we do not know the detail of $d_\mathrm{offset}$.
We take the following steps before performing the MCMC fitting.
\begin{itemize}
    \item The statistical uncertainties in the observations are smaller than the actual deviations from the model, and we thus introduce the contribution of systematic uncertainties possibly due to gain fluctuations, standing waves, and stray light.
    We use the MCMC method only with the detector offset as an input parameter and the design values for other parameters and determine an additional error so that the reduced chi-squared becomes 1.
    Although the leakage to another harmonic in the multiplier may also cause this systematic uncertainty, we do not see such a harmonic in the standing wave plot (see Appendix \ref{app:standing_wave}).
    Furthermore, the low transmittance of the HWP outside the observation frequency mitigates the effect of other harmonics in the multiplier.
    \item We use only data in the frequency range above 120.6 GHz for two reasons.
    One reason is that these points are away from the observation frequency.
    The other reason is that the value of $c/T$ in less than \SI{120}{GHz} is systematically smaller than the values expected from the design values of the HWP.
    Since fitted parameters of the HWP model deviate significantly from the design values if we include data from this region, we exclude data less than \SI{120.6}{GHz}.
\end{itemize}

\subsection{Results}
\label{sec4-4}
\begin{figure}[tbp]
    \centering
    \includegraphics[width=16cm]{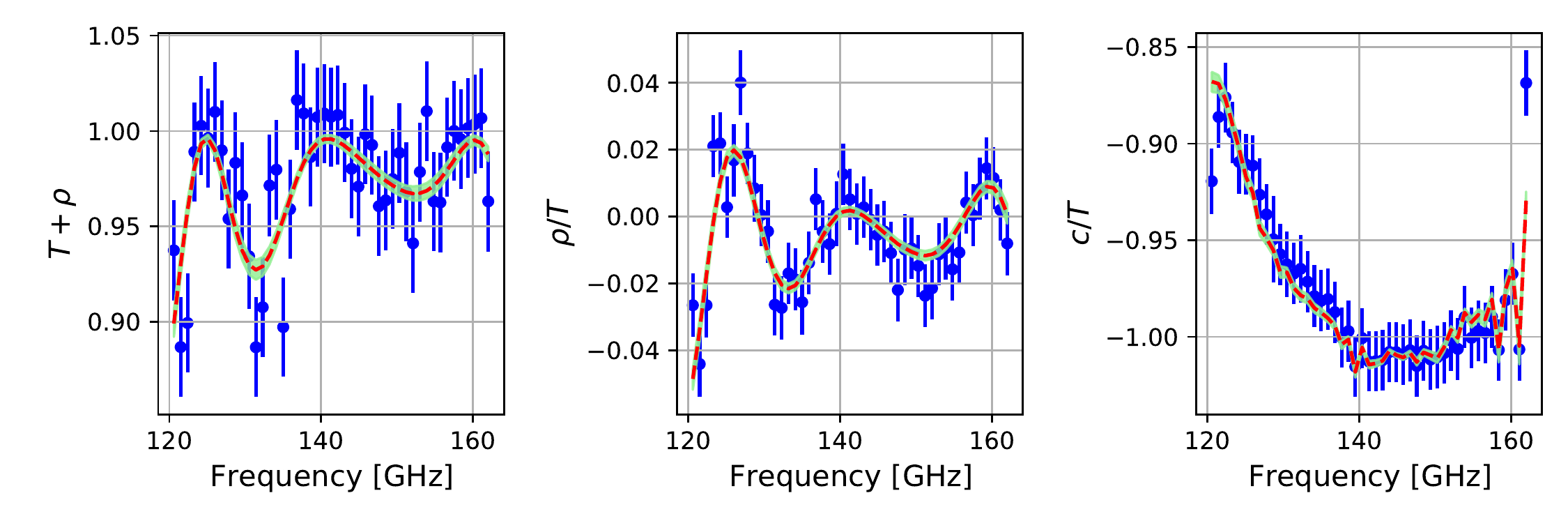}
    \caption{Calculated spectra of the Mueller matrix components and results of MCMC fitting. Blue points are measured points with error explained in section \ref{sec4-1}. The red dashed line and green band are the estimated mean value and one-sigma distribution respectively. Note that these data points and fitting lines include the effect of the offset of the lock-in amplifier.}
    \label{fig:mcmc_res1}
\end{figure}
The blue points in Figure~\ref{fig:mcmc_res1} show measured spectra of $T+\rho$, $\rho/T$, and $c/T$.
Figure~\ref{fig:mcmc_res1} also shows the spectra calculated with the HWP model with parameters from the MCMC.
The red dashed lines and green band indicate the converged average value and one-sigma band.
Note that these data points and fitting lines include the effect of the offset of the lock-in amplifier.
Table~\ref{tab:hwp_posterior} gives the parameters of the HWP model after the MCMC fitting.
In this table, we give the mean values of the posterior distribution.
A comparison with Table~\ref{tab:hwp_parameter} shows that the results are consistent with the fiducial values.
We also find that compared to the design value, the quantity of the uncertainty in the sapphire thickness becomes 1/20 and the quantity of the uncertainty in the refractive index difference of the sapphire becomes 1/3.
The accuracy of the model is improved.
Note that the posterior distribution of $d_\mathrm{offset}$ converges to $-0.009 \pm 0.001$.
\begin{table}[]
    \centering
    \caption{Parameters of the HWP model obtained by MCMC fitting. These values are calculated from the mean values of the posterior distributions and the uncertainties come from the samples' standard deviation. A comparison with Table~\ref{tab:hwp_parameter} shows that all parameters are consistent within the uncertainty given in the table. The uncertainty in the sapphire thickness and the uncertainty in the refractive index difference of the sapphire become smaller than the uncertainties in the design values.}
    \begin{tabular}{cccc}
        \hline \hline
         & thickness & refractive index & loss tangent \\
         \hline
         Sapphire (o-axis) & $3.086 \pm 0.005\, \mathrm{mm}$ & $3.065 \pm 0.002$ & $(1.11 \pm 0.89) \times 10^{-4}$ \\
         Sapphire (e-axis) & $3.086 \pm 0.005\, \mathrm{mm}$ & $3.404 \pm 0.002$ & $(0.51 \pm 0.50) \times 10^{-4}$ \\
         Duroid & $0.255 \pm 0.002$\, mm & $1.710 \pm 0.008$ & $(5.4 \pm 0.53) \times 10^{-4}$ \\
         LDPE & $0.039 \pm 0.002$\, mm & $1.514 \pm 0.010$ & $(4.3 \pm 4.7) \times 10^{-4}$ \\
         \hline \hline
    \end{tabular}
    \label{tab:hwp_posterior}
\end{table}

In this measurement, the uncertainties in the measured spectra are limited by the unknown systematic error.
The cause of this systematic error must be understood to improve the accuracy.
Additionally, the offset of the lock-in amplifier $d_\mathrm{offset}$ is estimated from the fitting as a nuisance parameter.
In future measurements, it would beneficial to measure the background signal to determine the offset.

Ideally, a vector network analyzer (VNA) would be used for the HWP characterization.
A VNA can measure the amplitude and phase of the electric field, allowing
the calculation of the Mueller matrix directly from Eq.~\eqref{eq:jones_to_mueller}.

\section{Leakage Estimation}
\label{sec5}
\begin{figure}
    \centering
    \includegraphics[width=16cm]{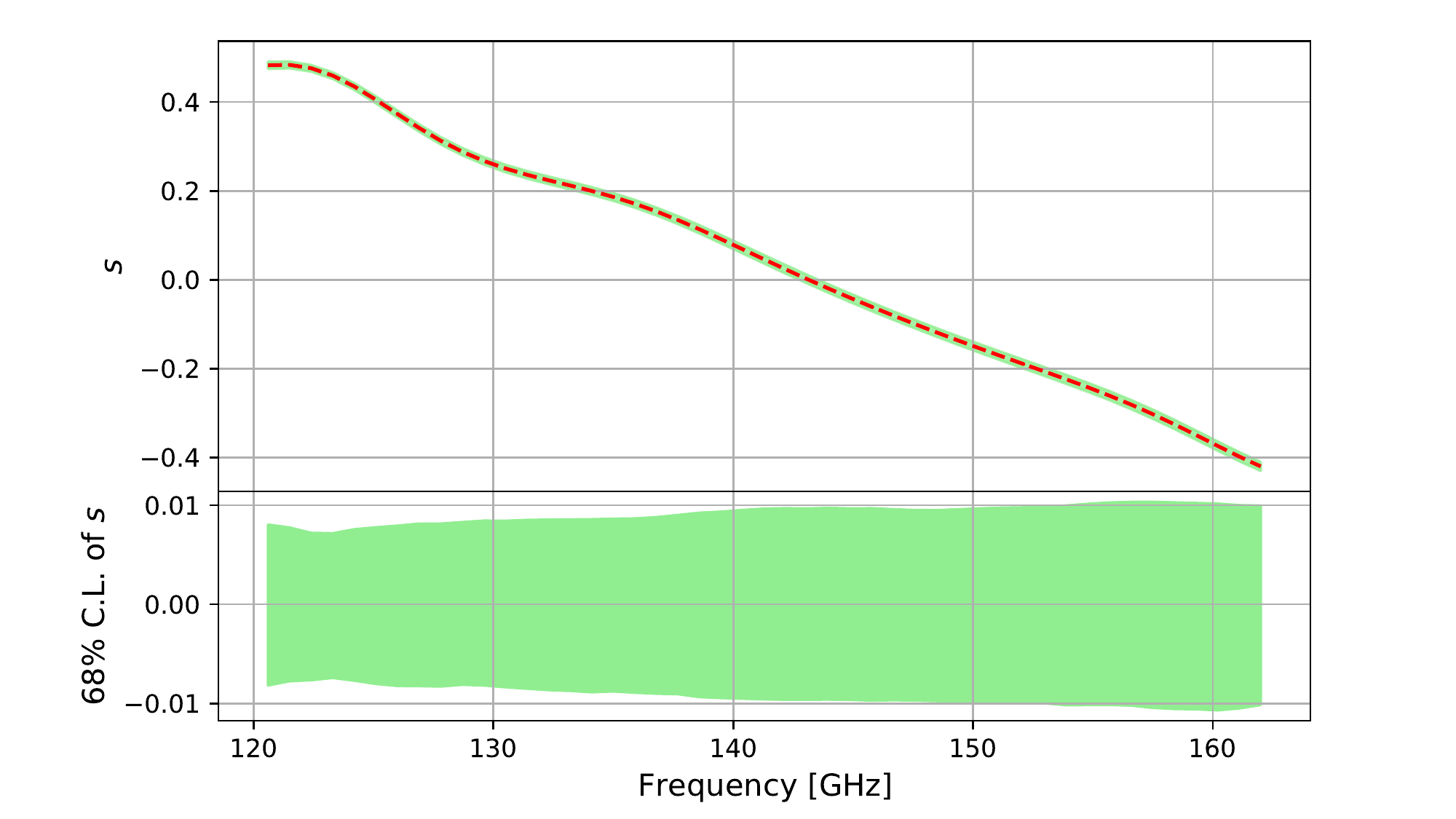}
    \caption{Spectra of the estimated $s$ parameter. The top panel shows the spectra of the mean $s$ value (red dash line) and the range of one sigma (green band). The bottom panel shows the one-sigma uncertainty of the $s$ parameter.}
    \label{fig:s_spec}
\end{figure}

We calculate the coupling between the CP and linear polarization, $s$, using the HWP model (see section~\ref{sec2-1}) and the parameters obtained from the samples of the MCMC fitting in section~\ref{sec4}.
Figure~\ref{fig:s_spec} shows the spectra of $s$.
The $s$ spectrum is close to zero at the observation frequency, which is given in Table~\ref{tab:wafer_bandpass}, and decreases at high frequency as expected from Eq.~\eqref{eq:mueller_matrix}.
The frequency at which $s$ is zero is approximately \SI{143}{GHz}.
This frequency is close to the central frequency of the \textsc{Polarbear} telescope (see Table~\ref{tab:wafer_bandpass}).
We show the uncertainty in this spectrum in the bottom panel.
The uncertainty is almost constant within the observation frequency range.

We calculate the band-averaged value, $\bar{s}$, using Eq.~\eqref{eq:band_averaged}.
We give the details of the bandpass of each wafer in Table~\ref{tab:wafer_bandpass}.
For the detector band-pass window function, $W(\nu)$, we use the wafer-averaged spectrum of the FTS measurement at the site (see section~\ref{sec3}).
The detector bandpass properties are summarized in Table~\ref{tab:wafer_bandpass} for each wafer.
We calculate $\bar{s}$ for each wafer because $\bar{s}$ is sensitive to the wafer-by-wafer variation of the band center.
The statistical uncertainty in $\bar{s}$ is calculated from the standard deviation of the samples of $\bar{s}$.
\begin{table}[]
    \centering
    \caption{Detector bandpass of each wafer. The average (AVG) and standard deviation (STD) of the band center and bandwidth are shown.
    These values are obtained from FTS measurement at the site\cite{FTScharacterization}.
    We also show the band edge and the uncertainty in the bandpass of each wafer.
    This uncertainty is the wafer-averaged square root of the sum of the squares of the statistical and systematic error.
    The statistical uncertainty comes from the in-wafer detector-to-detector sensitivity uncertainty.
    The systematic error is the same as in the FTS paper.
    }
    \begin{tabular}{c|C|C|C|C|c|C}
        \hline \hline
        & \multicolumn{2}{c}{Band Center (GHz)} & \multicolumn{2}{c}{Band Width (GHz)} & Band Edge (GHz)
& Uncertainty \\
        Wafer & AVG & STD & AVG & STD & \\
        \hline
        8.2.0 & 136.9 & 0.7 & 30.4 & 1.8 & 121.7 -- 152.1 & 0.007 \\
        9.4 & 146.9 & 0.5 & 32.8 & 1.6 & 130.5 -- 163.3 & 0.007 \\
        10.1 & 142.1 & 2.5 & 31.8 & 1.8 & 126.2 -- 158.0 & 0.010 \\
        10.2 & 143.5 & 0.5 & 32.6 & 1.1 & 127.2 -- 159.8 & 0.005 \\
        10.3 & 148.7 & 0.6 & 31.0 & 1.9 & 133.2 -- 164.2 & 0.007 \\
        10.4 & 144.0 & 0.5 & 32.2 & 1.2 & 123.9 -- 156.1 & 0.005 \\
        10.5 & 145.5 & 0.4 & 31.8 & 1.3 & 129.6 -- 161.4 & 0.006 \\
        \hline \hline
    \end{tabular}
    \label{tab:wafer_bandpass}
\end{table}

Unlike the case for the linear polarization of the CMB, there is a variation in the frequency dependence of the CMB CP.
Here, we consider four spectra of the CP of the CMB, namely the Rayleigh--Jeans (RJ) spectrum, CMB spectrum, the frequency dependence of the FC\cite{circ_prospects}, and the frequency dependence of the CP caused by C$\nu$B\cite{circ_CnB}:
\begin{align}
    \label{eq:spectra_rj}
    S_{\mathrm{RJ}}(\nu) &= S(\nu_0) \left( \frac{\nu}{\nu_0} \right)^2, \\
    \label{eq:spectra_cmb}
    S_{\mathrm{CMB}}(\nu) &= S(\nu_0) \left( \frac{\nu}{\nu_0} \right)^4 
    \frac{\exp(h \nu/ k_B T) / \exp(h \nu_0/ k_B T)}{(\exp(h \nu / k_B T)-1)^2 / (\exp(h \nu_0 / k_B T)-1)^2}, \\
    \label{eq:spectra_f3}
    S_{\mathrm{FC}}(\nu) &= S_{\mathrm{CMB}}(\nu) \left( \frac{\nu_0}{\nu} \right)^3, \\
    \label{eq:spectra_f1}
    S_{\mathrm{C\nu B}}(\nu) &= S_{\mathrm{CMB}}(\nu) \frac{\nu_0}{\nu},
\end{align}
where $S(\nu_0)$ is the amplitude of the signal at the pivot frequency $\nu_0$, and $h, k_B, T$ are the Planck constant, Boltzmann constant, and temperature of the CMB (2.725~K)\cite{CMB_temp}.
We also consider the spectrum of CP due to the atmospheric Zeeman emission\cite{atmos_circular_2003, class_atm}.
The Zeeman emission is a signal produced by the splitting of the energy levels of the oxygen molecules in the atmosphere by the Earth's magnetic field.
This signal is expected in that the low-frequency side of the split level at \SI{118.75}{GHz} is circularly polarized clockwise and the high-frequency side is circularly polarized counterclockwise.

Table~\ref{tab:s_results2} presents the band- and spectral-dependence of the band-averaged $s$ value.
The values of wafer 8.2.0 are larger than those of other wafers because the central frequency of wafer 8.2.0 is lower than that of the other wafers.
This relation of $\bar{s}$ among wafers is independent of the source spectrum.

Regarding the source spectrum dependence, the values of the CMB spectrum are approximately $0.02$ larger than those of the RJ spectrum.
The maximum absolute value of our estimates is almost the same as that of the SPIDER HWP $s$ parameters\cite{circ_SPIDER_result} (0.149 in this paper versus 0.154 in SPIDER), which are also band-averaged using the CMB source spectrum and their bandpass.
However, the minimum value in this paper is larger than that in SPIDER (0.007 versus 0.003).
We also compare the uncertainty in the $s$ parameter.
The uncertainty in the $s$ parameter of the SPIDER HWP is approximately 0.041 whereas
the uncertainty in the $s$ parameter obtained from the design values of \textsc{Polarbear} HWP (Table~\ref{tab:hwp_parameter}) is approximately 0.1. 
In contrast to these results, the uncertainty in our estimated $s$ parameter is approximately 0.009.
The method described in this paper thus reduces the uncertainty in the HWP model.

Meanwhile, the values of the FC spectrum, C$\nu$B spectrum, and Zeeman spectrum are different.
Although some signs of band-integrated $s$ values are reversed, the set of absolute values of the FC spectrum and C$\nu$B spectrum are almost the same as those of the RJ spectrum and CMB spectrum.
The band-integrated $s$ values are larger for the Zeeman spectrum than for the other spectra.
The CP signal of the Zeeman emission is expected to be approximately 61~$\mu$K in the \textsc{Polarbear} frequency band.
Even with the suppression by $\bar{s}$, the apparent signal is approximately 31~$\mu$K, which is above the noise level of \textsc{Polarbear} ($\mathrm{NET_{array}} = 
\SI{23}{\micro\kelvin\sqrt{s}}$
)\cite{large_patch}.
\begin{table}[]
    \centering
    \caption{Band-averaged $s$ values for various spectra of sources. In this paper, we assume Rayleigh--Jeans spectrum (RJ), CMB spectrum (CMB), spectrum of the CP due to Faraday Conversion (FC), spectrum of the CP due to the cosmic neutrino background (C$\nu$B), and atmospheric Zeeman emission (Zeeman). The difference in the values between the RJ and CMB is smaller than the differences between other spectra. Meanwhile, the values of FC and C$\nu$B differ largely from those of RJ. Moreover, the values of the atmospheric Zeeman emission are larger than those of the other spectra.}
    \begin{tabular}{cccccc}
        \hline \hline
        Wafer & RJ & CMB & FC & C$\nu$B & Zeeman \\
        \hline
        8.2.0 & 0.133 $\pm$ 0.009 & 0.149 $\pm$ 0.009 & 0.207 $\pm$ 0.008 & 0.168 $\pm$ 0.009 & 0.437 $\pm$ 0.011 \\
        9.4 & -0.075 $\pm$ 0.009 & -0.055 $\pm$ 0.009 & -0.014 $\pm$0.009 & -0.033 $\pm$ 0.009 & 0.366 $\pm$ 0.009 \\
        10.1 & 0.024 $\pm$ 0.009 & 0.046 $\pm$ 0.009 & 0.123 $\pm$ 0.009 & 0.070 $\pm$ 0.009 & 0.421 $\pm$ 0.008 \\
        10.2 & -0.002 $\pm$ 0.009 & 0.018 $\pm$ 0.009 & 0.089 $\pm$ 0.009 & 0.041 $\pm$ 0.009 & 0.396 $\pm$ 0.009 \\
        10.3 & -0.113 $\pm$ 0.010 & -0.094 $\pm$ 0.010 & -0.029 $\pm$ 0.009 & -0.073 $\pm$ 0.009 & 0.322 $\pm$ 0.009 \\
        10.4 & -0.013 $\pm$ 0.009 & 0.007 $\pm$ 0.009 &0.076 $\pm$ 0.009 & 0.029 $\pm$ 0.009 & 0.389 $\pm$ 0.008 \\
        10.5 & -0.045 $\pm$ 0.009 & -0.025 $\pm$ 0.009 & 0.041 $\pm$ 0.009 & -0.004 $\pm$ 0.009 & 0.367 $\pm$ 0.009 \\
        \hline \hline
    \end{tabular}
    \label{tab:s_results2}
\end{table}

\section{Discussion}
\label{sec6}

\subsection{Systematic Uncertainty}
\label{sec6-1}
We estimate the systematic uncertainties in the band-averaged $s$ parameter of the HWP under the conditions of actual operation in the telescope.
We consider the uncertainty in the bandpass dependence and the non-vertical incident light.

We first consider the uncertainty in the bandpass dependence.
There is uncertainty in the bandpass measurement by the FTS at the site.
This comes from both statistical and systematic uncertainty.
The statistical uncertainty comes from the in-wafer detector-to-detector sensitivity uncertainty.
The systematic uncertainty comes from both time-variability in the measurement and from errors in alignment of the FTS, which is same as the FTS paper.
We calculate the uncertainty in the bandpass of each wafer from the data in the sensitive frequency region as shown in Table~\ref{tab:wafer_bandpass}.
We then calculate $\bar{s}$ 5000 times using random realizations of the detector bandpass with the uncertainty evaluated above.
We take the standard deviation of this distribution as the systematic uncertainty due to the uncertainty in the detector bandpass.

We next consider the non-vertical incident light.
In the HWP model used in section~\ref{sec5}, we assume that light is vertically incident. 
However, not all light is vertically incident in the setup of the telescope.
The \textsc{Polarbear} HWP is placed at the prime focus, which is between the primary and secondary mirrors, of the Huan Tran Telescope.
The light between these mirrors is once focused at the prime focus and spreads again, and the incident angle of the light incident on the HWP thus increases as the light deviates from the center of the optical path.
The non-vertical incident light will change the optical path in the HWP and thus affect the estimate of $\bar{s}$.
The maximum value of the incident angle is $16^{\circ}$ at the half width at half maximum from the geometry.
We thus calculate conservatively how $\bar{s}$ varies when the incident light is tilted at $16^{\circ}$.
In this calculation, we rotate the HWP with the tilted incident light and extract the second harmonic in the simulation.

Table \ref{tab:s_results1} gives the systematic uncertainties in the band-averaged $s$ value.
Here, we assume that the source spectrum is the RJ spectrum.
The columns from the left show the wafer name, band-averaged $s$ values, statistical uncertainty in the band-averaged $s$ value, systematic error of the uncertainty in the detector bandpass, and systematic error in the non-vertical incident light.
We find that these systematic uncertainties are smaller than the statistical uncertainty.
\begin{table}[]
    \centering
    \caption{Estimated $\bar{s}$ and uncertainties for each wafer.
    The first and second columns from the left give the average values and standard deviation of the MCMC fitting explained in section~\ref{sec5}.
    The third column gives the systematic error due to the uncertainty in the detector bandpass.
    The fourth column gives the systematic error due to the non-vertical incident light.
    All values in this table are calculated by assuming the Rayleigh--Jeans spectrum.
    These systematic uncertainties are smaller than the standard deviation of the MCMC fitting.}
    \begin{tabular}{cCCCC}
        \hline \hline
        Wafer & AVG & STD & bandpass & non-vertical \\
        \hline
        8.2.0 & 0.133 & 0.009 & 0.001 & 0.001 \\
        9.4 & -0.075 & 0.009 & 0.001 & 0.002 \\
        10.1 & -0.024 & 0.009 & 0.001 & 0.002 \\
        10.2 & 0.002 & 0.009 & 0.001 & 0.003 \\
        10.3 & -0.113 & 0.010 & 0.001 & 0.002 \\
        10.4 & -0.013 & 0.009 & 0.001 & 0.003 \\
        10.5 & -0.045 & 0.009 & 0.001 & 0.003 \\
        \hline \hline
    \end{tabular}
    \label{tab:s_results1}
\end{table}

\subsection{Possibility of cross-checking using atmospheric CP}
\label{sec6-2}
We next consider a method of cross-checking the above result with the values obtained from observation. 
The atmospheric Zeeman emission is a possible CP source with which to measure $\bar{s}$.
As described in section~\ref{sec5}, the atmospheric Zeeman emission is a bright CP source and is expected to be observable with the \textsc{Polarbear} detector.
Thus, we might be able to separate the CP signal from the second harmonic signal using the method described in section~\ref{sec2-2} and estimate the leakage of the HWP by comparing the observed CP signal with the theoretical value.

Note that because the coordinate of the atmospheric CP is fixed to the ground, the signal of the atmospheric CP may be degenerated with the ground pickup signal.
The difference in spectra can be used to distinguish the atmospheric CP signal.
The Zeeman emission has a peak at \SI{118.75}{GHz} and this results in a temperature difference of approximately 100 $\mu$K at maximum between wafers due to slight differences in frequency characteristics.
If the spectrum of the ground pickup signal is the RJ spectrum, we can distinguish the Zeeman emission from the difference in the observed temperature between wafers.

\subsection{Prospects of CP Measurement}
\label{sec6-3}
Table IV in section V shows that $\bar{s}$ is nonzero in most cases.
This suggests the possibility to probe the CP using \textsc{Polarbear} data.
We roughly estimate the sensitivity of $\ell (\ell+1) C_\ell^{VV} /(2 \pi)$ by scaling the uncertainties of $B$-mode observation\cite{reanalysis} by the mean value of the $\bar{s}^{-2}$ in Table~\ref{tab:s_results2} of the CMB spectrum.
As a result, the upper limit is approximately \SI{30}{(\micro K)^2} at $\ell \sim 300$ in \textsc{Polarbear}.
This value is less than 1/8 compared to the sensitivity of measurements made by SPIDER.
\section{Summary}
\label{sec7}
We evaluated the HWP used at \textsc{Polarbear}, including the leakage between linear polarization and CP.
We constructed a HWP model from data recorded at the laboratory in 2014 and estimated the leakage between the CP and linear polarization.
This model well explained the measured spectrum of the Mueller matrix components, and the uncertainty in the parameters of the HWP was at maximum 1/20th the design value.
We thus found that the absolute value of the band-averaged leakage from the CP obtained using the HWP, $\bar{s}$, ranged from $0.133$ to $0.002$ at each wafer, and the statistical uncertainty in $\bar{s}$ was approximately $0.010$ for each wafer in the case of the RJ spectrum.
This means that almost all detectors on each wafer are capable of measuring the CP.
We also considered four other spectra.
The value of $\bar{s}$ was nonzero in most cases.
In particular, $\bar{s}$ was larger for the Zeeman spectrum than for the other spectra.
We also estimated the systematic uncertainties in $\bar{s}$.
In this paper, we considered the uncertainties in the detector bandpass and non-vertical incident light
and found that these systematic uncertainties were smaller than the statistical error.
Finally, we indicated the method to verify this result using the atmospheric CP signal
and presented prospects of making angular power spectrum measurements of CP anisotropy.
Although the projected sensitivity from the \textsc{Polarbear} HWP is not sufficient to detect the estimated CP signal due to the population III\cite{circ_popIII} which is approximately less than \SI{10e-12}{(\micro K)^2} at \SI{150}{GHz}, it is better than the currently given upper limits at small angular scales.


%
%

%

\begin{acknowledgments}
We acknowledge M. J. Myers for creating the HWP, and Y. Inoue and H. Yamaguchi for setup the laboratory measurement system.
MH acknowledges support from the World Premier International Research Center Initiative (WPI) of MEXT and the JSPS KAKENHI grant No. JP22H04945.
ST acknowledges support from the JSPS KAKENHI grant Nos. JP14J01662 and JP18J02133.
HN achnowledges support from the JSPS KAKENHI grant No. JP17K18785.
This work was supported by the JSPS Core-to-Core Program.
We thank Edanz (https://jp.edanz.com/ac) for editing a draft of this manuscript.
\end{acknowledgments}

\appendix

\section{Standing wave removal}
\label{app:standing_wave}
To remove the standing wave effect, we fit the data at each frequency and rotation angle with the following equation,
\begin{equation}
    \label{eq:standing_wave_rm}
    d(z) = A \sin(Bz+C)+Dz+E,
\end{equation}
where $z$ is the detector position and $d$ is the amplitude at each detector position.
$A, B, C, D$, and $E$ are free parameters: the amplitude of the standing wave, the wave number, the phase, the drift, and the constant component respectively.
\begin{figure}
    \centering
    \includegraphics[width=16cm]{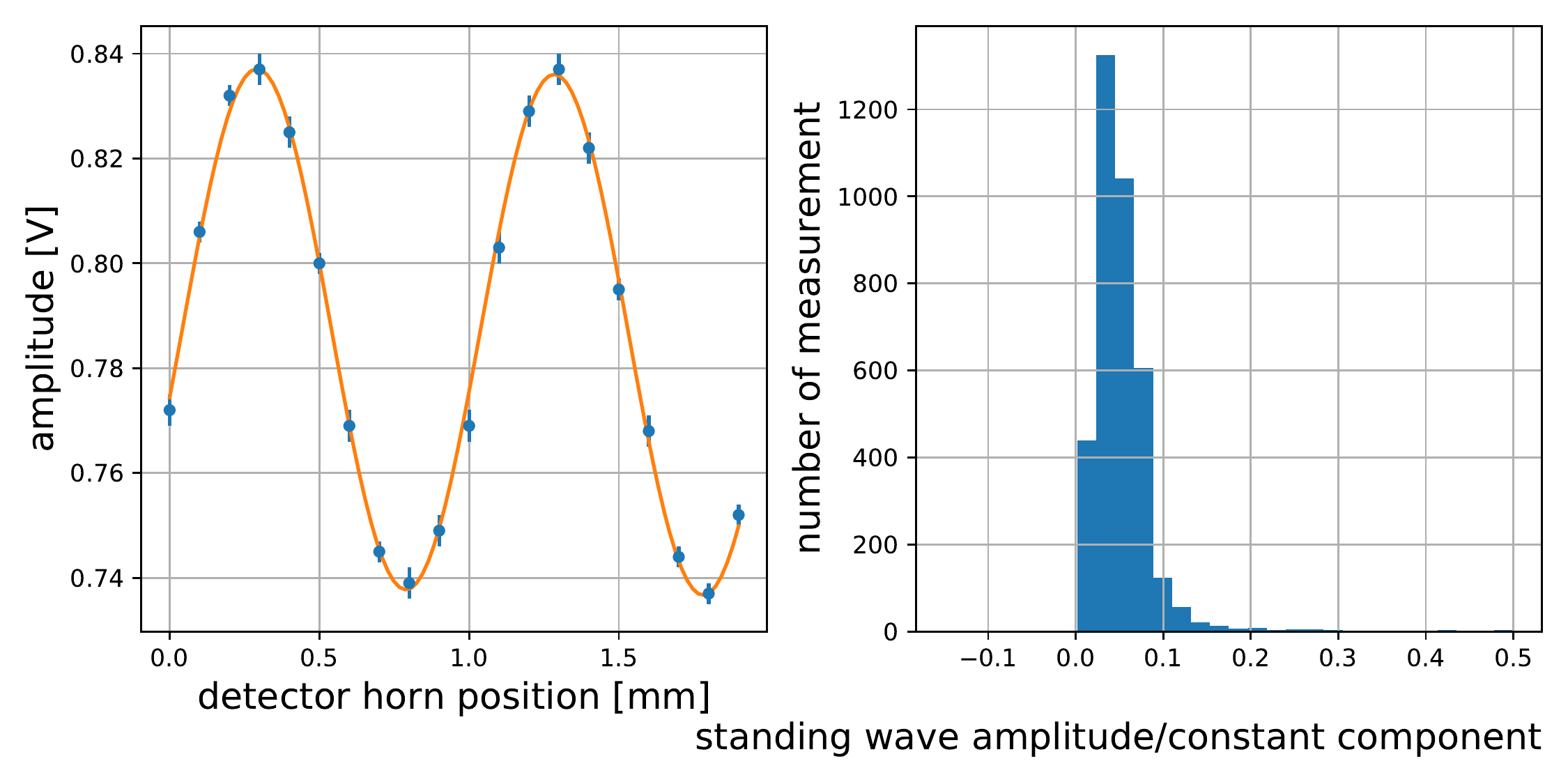}
    \caption{(Left) A sample of standing wave signal (blue points) and fitting result (orange line).
    (Right) Distribution of the amplitude of the standing wave over the constant component.
    We find that these values are less than 0.1 for most measurements.}
    \label{fig:standing_wage}
\end{figure}
The left panel of Figure \ref{fig:standing_wage} shows a sample of the standing wave signal and its fitting result.
Then we extract the $E$ value as the signal for this frequency and rotation angle.
The right panel of Figure \ref{fig:standing_wage} shows the distribution of the ratio of the amplitude of the standing wave $A$ to the constant component $E$.
These values are less than 0.1 for most of the measurements, thus the effect of the standing wave is not significant.

\section*{author declarations}

\subsection*{Conflict of Interest}
The authors have no conflicts to disclose.

\subsection*{Author Contributions}
\textbf{T. Fujino}:Conceptualization (equal); Formal Analysis (lead); Writing/Original Draft Preparation (lead).
\textbf{S. Takakura}: Conceptualization (equal); Investigation (lead); Writing/Review \& Editing (lead).
\textbf{Y. Chinone}: Writing/Review \& Editing (supporting).
\textbf{M. Hasegawa}: Writing/Review \& Editing (equal).
\textbf{M. Hazumi}: Writing/Review \& Editing (supporting).
\textbf{N. Katayama}: Writing/Review \& Editing (supporting).
\textbf{A. T. Lee}: Writing/Review \& Editing (supporting).
\textbf{T. Matsumura}: Writing/Review \& Editing (equal).
\textbf{Y. Minami}: Writing/Review \& Editing (equal).
\textbf{H. Nishino}: Writing/Review \& Editing (supporting).

\bibliography{aimtemplate}

\end{document}